\documentclass[letterpaper,aps,prb,twocolumn,showpacs,preprintnumbers,superscriptaddress]{revtex4-1}

\usepackage{graphicx}
\usepackage{dcolumn}
\usepackage{bm}
\usepackage{amsmath}
\usepackage{epsfig}

\begin{document}

\title{Strong anisotropy in surface kinetic roughening: analysis and experiments}

\author{Edoardo Vivo}
\affiliation{Departamento de
Matem\'{a}ticas and Grupo Interdisciplinar de Sistemas Complejos
(GISC), Universidad Carlos III de Madrid, Avenida de la
Universidad 30, E-28911 Legan\'{e}s, Spain}
\author{Matteo Nicoli}
\affiliation{Physique de la Mati\`ere Condens\'ee, \'Ecole Polytechnique - CNRS,
91128 Palaiseau, France}
\author{Martin Engler}
\affiliation{II. Physikalisches Institut, Universit\"at zu K\"oln, Z\"ulpicherstr.\ 77,
D-50937 K\"oln, Germany}
\author{Thomas Michely}
\affiliation{II. Physikalisches Institut, Universit\"at zu K\"oln, Z\"ulpicherstr.\ 77,
D-50937 K\"oln, Germany}
\author{Luis V\'azquez}
\affiliation{Instituto de Ciencia de Materiales de Madrid, Consejo Superior de Investigaciones \\Cient\'{\i}ficas,
E-28049 Madrid, Spain}
\author{Rodolfo Cuerno}
\affiliation{Departamento de
Matem\'{a}ticas and Grupo Interdisciplinar de Sistemas Complejos
(GISC), Universidad Carlos III de Madrid, Avenida de la
Universidad 30, E-28911 Legan\'{e}s, Spain}

\date{\today}

\begin{abstract}
We report an experimental assessment of surface kinetic roughening properties that are anisotropic in space.
Working for two specific instances of silicon surfaces irradiated by ion-beam sputtering under diverse
conditions (with and without concurrent metallic impurity codeposition), we verify the predictions and
consistency of a recently proposed scaling Ansatz for surface observables like the two-dimensional (2D)
height Power Spectral Density (PSD). In contrast with other formulations, this Ansatz is naturally tailored
to the study of two-dimensional surfaces, and allows us to readily explore the implications of anisotropic scaling for
other observables, such as real-space correlation functions and PSD functions for 1D profiles of the surface.
Our results confirm that there are indeed actual experimental systems whose kinetic roughening is strongly anisotropic,
as consistently described by this scaling analysis. In the light of our work, some types of experimental measurements
are seen to be more affected by issues like finite space resolution effects, etc.\ that may hinder a clear-cut assessment of strongly anisotropic scaling in the present and other practical contexts.
\end{abstract}

\pacs{
68.35.Ct, 
05.40.-a, 
79.20.Rf, 
}

\maketitle


Surface kinetic roughening, namely, the lack of characteristic length and time scales in the fluctuations of the values
of an interface topography, is quite a pervasive phenomenon, from hard to soft condensed matter physics systems. It occurs for instance in thin-film production,\cite{michely:book} crack formation,\cite{bouchaud:1997} fluid dynamics,\cite{mccomb:book} or growth of cellular aggregates.\cite{barabasi:book} Naturally, in each specific context, the meaning and even the dimensionality of the relevant surface or interface needs to be unambiguously defined. In our case,
we will consider as reference systems proper two-dimensional surfaces of, e.g., thin films, produced by
chemical or vapor deposition techniques, epitaxy, ion-beam erosion, etc.\cite{zhao:book}

In spite of the large efforts devoted to the understanding of kinetic roughening, a number of issues remain poorly understood. This lack of complete knowledge is perhaps one important reason for the relatively moderate experimental impact that this general phenomenon has met, see [\onlinecite{cuerno:2007}] and references therein. One such aspect is the phenomenon of anisotropic scaling, namely, when the properties associated with scale invariance in the fluctuations
of the surface height change with the substrate direction one is looking along. Technically, this is reflected in the fact that the ensuing power-law behavior of observables in the system\cite{barabasi:book} are characterized by direction-dependent critical exponents.

Perhaps the most obvious class of systems in which topographical anisotropies occur correspond to those in which morphological instabilities induce different typical length-scales along different directions. For instance, for solid targets eroded by ion-beam sputtering (IBS), an oblique ion flux induces ripple formation, \cite{chan:2007,munoz-garcia:2009} breaking the symmetry between the substrate directions along the ripples and perpendicular to them. Note, however, that for such pattern-forming systems scale invariance does not occur at length scales that are comparable with the typical wavelength of the pattern. Nevertheless, kinetic roughening can still occur simultaneously at length scales that are, rather, well separated from such a preferred scale. Examples are known in, e.g., the context of the Kuramoto-Sivashinsky (KS) equation,\cite{nicoli:2010} frequently used to model IBS systems;\cite{munoz-garcia:2009} under such conditions, it is just natural to expect the ensuing kinetic roughening properties to take on an anisotropic form.\cite{keller:2009}

For morphologically stable systems, one can conceive of anisotropic scaling behavior due to, e.g., physical anisotropies in the external flux that is driving the surface dynamics. This is the case, for instance, for macroscopic geological systems, in which an external matter flux has a direction that is preferred, e.g., by gravity on a tilted landscape.\cite{pastor-satorras:1998} Anisotropies can also exist in the relaxation mechanisms that are intrinsic to the system: e.g., anisotropies in surface tension in phenomena of crystal growth from a melt,\cite{davis:book} or in barriers to relaxation by surface diffusion, even for a negligible (unstable) Ehrlich-Schwoebel effect, in the case of epitaxial growth.\cite{michely:book}

Given the above, remarkably few reports exist in which anisotropic kinetic roughening has been unambiguously assessed
(see, e.g., Ref.\ \onlinecite{cuerno:2004} and references therein), especially as taking place in the steady state of the system, which is what we will call strong anisotropy (SA). From the theoretical point of view, this is the most interesting manifestation of anisotropic behavior, since it corresponds to the behavior of a system in the thermodynamic (or, for out-of-equilibrium systems, more properly ``hydrodynamic'') limit. Perhaps one of the complications in the identification of SA is the fact that this type of behavior had been previously encoded in a scaling Ansatz \cite{schmittmann:2006} that originates in the study of critical dynamics of equilibrium statistical-mechanical systems.\cite{henkel:book_v1} Such a formulation is quite powerful from the theoretical
point of view (enabling analysis of scaling properties for arbitrary substrate dimension, etc.) but is not
particularly natural for the characterization of actual two-dimensional surfaces. In this way, the experimental thin-film or surface-science community does not have a clear guiding principle that allows it to identify correctly the behavior that should be expected for each observable under conditions for strong anisotropy. Comparison with theoretical models is moreover hampered. A clear-cut picture is particularly required in situations in which finite size or analogous effects may interfere with an underlying scaling behavior.

Some formulations of SA  that adapt better to the thin film context are available,\cite{zhao:book} but have been tested on simplified models only, \cite{zhao:1998} so that their generality has not been checked, nor has a systematic study been done on the relation between different observables within such a framework. Specifically, given the widespread current use of scanning probe microscopies [like atomic force microscopy (AFM) or scanning tunneling microscopy (STM)] and of X-ray diffraction techniques, it is important to understand the relationship between the scaling behaviors of the two-dimensional Power Spectral Density (PSD) of the surface height and the PSD of one-dimensional cuts along different substrate directions, and their relation with real-space height correlation functions (see below for precise definitions). This is due to the fact that many experimental setups are often designed to
measure either the former (X-rays), or the latter (AFM/STM).\cite{zhao:book} Therefore, a fully consistent analysis and a quantitative comparison with theoretical models cannot be done unless this relationship is fully clarified.

The purpose of this work is twofold: First, considering a recently proposed anisotropic scaling Ansatz that has been
tested against linear and nonlinear models,\cite{us_theor} we elucidate the behavior that is expected for
one- and two-dimensional PSDs in a shape that is natural to thin-film analysis techniques, and the relation with the behavior of real-space height correlation functions. Note that anisotropic scaling is a physical property that
experimental systems may or may not have, independently in principle of comparison with theoretical models (which, in turn, may or may not have it on their own).

Once this framework for analysis is clarified, we address the experimental observation of SA in two cases. They both correspond to experiments of surface erosion of silicon targets by ion-beam sputtering. But one of them  is within a morphologically stable condition in which metallic contaminants are suppressed, while a second one corresponds to erosion with concurrent deposition of metallic contaminants leading to a morphological instability. At any rate, the occurrence or not of the instability will not play any role in our discussion, as we will focus on length scales at which scale invariance holds in each case. Thus, we are able to discuss in detail various aspects on the occurrence of SA that may have hampered a more clear assessment of this property in the past. Based on these results, we propose forms of analysis that seem less susceptible for complications and ambiguities.

This paper is organized as follows. In Sec.\ \ref{sec:asa}, we discuss the anisotropic scaling Ansatz that will be
employed for data analysis in later parts of the work. We describe its implications for height measurements that are performed both in real and in Fourier space, for the full two-dimensional surface, and/or for line profiles of it. Section \ref{sec:exp} is then devoted to the application of this scaling Ansatz to the description of our experimental data. Finally, we present a discussion of these results from a general point of view and conclude in Sec.\ \ref{sec:disc}.
Supplemental Material is provided \cite{suppl} that contains morphological data collected by AFM and STM as well as some auxiliary plots, that has been employed in order to produce the graphs and the discussion provided in Section \ref{sec:exp}.

\section{Anisotropic scaling Ansatz}
\label{sec:asa}

%


Consider a surface whose height above a reference plane is given by a height function $h(\mathbf{r})$, where $\mathbf{r}=(x,y)$ is a point on a reference substrate plane. If the surface morphology is disordered,\cite{barabasi:book} a quantity that conveniently characterizes the fluctuations of the height around its average value is the height-difference correlation function
\begin{equation}
G(r) = \langle [h(\mathbf{r}+\mathbf{r}_0)-h(\mathbf{r}_0)]^2 \rangle ,
\label{G}
\end{equation}
which is usually a function of relative distance, $r=|\mathbf{r}|$, only. Here, brackets denote average over, say, different experiments carried out under the exact same conditions, and $\mathbf{r}_0=(x_0,y_0)$ is an arbitrary position on the substrate. Isotropic kinetic roughening corresponds to a situation in which this correlation function grows as a power law with distance, $G(r) \sim r^{2 \alpha}$, where $\alpha$ is the so-called roughness exponent.\cite{barabasi:book} This exponent has a direct relation with, e.g., the fractal dimension of the surface, thus characterizing its asperity or roughness properties. Note, the term ``kinetic roughening'' suggests in particular that our surface is evolving in time. Nevertheless, unless stated otherwise, we assume in this section that a steady state has been reached in which morphological properties do not change on average.

For an anisotropic surface, the value of the roughness exponent changes with the direction along the substrate plane, thus one has $G_x(x)\sim x^{2\alpha_x}$ and $G_y(y)\sim y^{2\alpha_y}$, where
\begin{equation}
G_x(x) = \langle \left[h(x_0+x,y_0) - h(x_0,y_0)\right]^2\rangle,
\end{equation}
and $G_y(y)$ is defined analogously. The system is said to display \emph{Strong Anisotropy} (SA) if indeed $\alpha_x \neq \alpha_y$, whereas we talk about \emph{Weak Anisotropy} (WA), when the steady state of the system is actually \emph{isotropic}.

In our analysis we focus on the power spectral density (PSD) of the surface height,
\begin{equation}
S(\mathbf{k}) = \langle |\tilde{h}(\mathbf{k})|^2 \rangle ,
\label{psd}
\end{equation}
where $\tilde{h}(\mathbf{k})$ is the space Fourier transform of $h(\mathbf{r}) - \bar{h}$, with $\bar{h}$ being the space
average of the height, and $\mathbf{k}=(k_x,k_y)$ is the wave vector. Our aim is to formulate a scaling form for $S(\mathbf{k})$ that is ``typical'' of strongly anisotropic systems, and study how such a behavior reflects in other observable functions related with the PSD and/or the height-difference correlation function. For a strongly anisotropic system, we postulate \cite{us_theor} that the stationary PSD scales with wave-vector components $k_x$ and $k_y$ as
\begin{equation}
\label{psd2d}
S(k_x,k_y) \sim \left(k_x^{2\tilde{\alpha}_x} + \nu k_y^{2\tilde{\alpha}_y}\right)^{-1},
\end{equation}
where we refer to $\tilde{\alpha}_x$ and $\tilde{\alpha}_y$ as roughness exponents in momentum space, and $\nu$ is a mere constant. Note that the ``asymptotic to'' ($\sim$) sign expresses an ``equality'' that holds, up to numerical constants, in an appropriate small-$k=|\mathbf{k}|$ (equivalently, large-$r$) approximation, as  customary in critical behavior.\cite{henkel:book_v1} Note also that, in principle, the behavior expressed by Eq.\ \eqref{psd2d} can be directly checked from, say, AFM or STM data treated in Fourier space. Moreover, the PSD function $S(\mathbf{k})$ does behave as described by Eq.\ \eqref{psd2d} for a number of representative linear and nonlinear models of anisotropic surfaces.\cite{us_theor} As an example, consider a surface that relaxes by surface tension along the $x$ direction
and by surface-diffusion along the $y$ direction. The simplest evolution equation for the surface height that incorporates fluctuations in, say, an external flux of adatoms reads\cite{zhao:1998,us_theor}
\begin{equation}
\label{eq_2-4}
\partial_t h = \nu_x \partial_x^2 h - \nu_y \partial_y^4 h + \eta(x,y,t),
\end{equation}
where $\nu_{x,y}$ are positive constants related with surface tension and surface diffusivity,\cite{michely:book,barabasi:book} and $\eta$ is a noise term. For this linear model, it can be readily shown that, in the long time $t\to\infty$ limit, the PSD function takes exactly the form given by Eq.\ \eqref{psd2d} with $\tilde{\alpha}_x = 1$ and $\tilde{\alpha}_y = 2$; more elaborate models also comply with the same functional form, with different values for the $\tilde{\alpha}_{x,y}$ exponents, see Ref.\ \onlinecite{us_theor} and Sec.\ \ref{sec:exp} below.

In many experimental systems, the two-dimensional PSD is too noisy to get a reliable estimation of the critical exponents, and a different approach is preferred. A simple way to increase the signal-to-noise ratio is to take one-dimensional profiles of the surface topography $h(x,y)$ along the $x$ (or the $y$) direction and Fourier transform them.
Given a point $y_0$ (or, equivalently, $x_0$) one can show that the continuum limit of the PSD of this one-dimensional cut  is obtained by integrating the two-dimensional PSD over the $k_y$ direction~\cite{us_theor}
\begin{equation}
\label{psd1d}
S_x(k_x) = \dfrac{1}{\pi}\int_0^{\infty} dk_y\, S(k_x,k_y).
\end{equation}
Obviously, the same relation holds for the PSD of the one-dimensional cut along the $y$ direction,
i.e.\ $S_y(k_y)$. Inserting Eq.\ \eqref{psd2d} into Eq.\ \eqref{psd1d}, we obtain scaling laws for
$S_x$ and $S_y$
\begin{eqnarray}
\label{eq:psd1d_scaling}
&&S_x(k_x)\sim k_x^{-(2\tilde{\alpha}_x-\zeta)} , \\[5pt]
&&S_y(k_y)\sim k_y^{-(2\tilde{\alpha}_y-1/\zeta)} . \label{eq:psd1d_scaling2}
\end{eqnarray}
In these relations, we have introduced the exponent ratio \mbox{$\zeta = \tilde{\alpha}_x / \tilde{\alpha}_y$} that measures the degree of anisotropy of the surface, and will be referred to as an {\em anisotropy exponent}. We say that the system displays SA whenever $\zeta \neq 1$, while $\zeta = 1$ for isotropic systems. As shown below,\cite{keller:2009} Eqs.\ \eqref{eq:psd1d_scaling} and \eqref{eq:psd1d_scaling2} are equivalent to
\begin{eqnarray}
\label{eq:psd1d_scaling_b}
&&S_x(k_x)\sim k_x^{-(2\alpha_x+1)} , \\[5pt]
&&S_y(k_y)\sim k_y^{-(2\alpha_y+1)} , \label{eq:psd1d_scaling_b2}
\end{eqnarray}
that provide the natural generalization to the SA case of the scaling behavior of the PSD of 1D cuts of the surface in the isotropic case, in which $\alpha_x=\alpha_y=\alpha$ and $S_{x,y} \sim k_{x,y}^{-(2\alpha+1)}$.\cite{hansen:2001}
Still, we have to prove that the anisotropy condition $\zeta\neq 1$ indeed corresponds to
$\alpha_x \neq \alpha_y$, for which we need to relate these real-space exponents with their momentum-space counterparts, $\tilde{\alpha}_{x,y}$. The required relation is readily obtained from the definition of the height-difference correlation function along one direction, e.g., along the $x$ coordinate, $G_x(x)$ above. To increase the precision of the estimate of $G_x$, this function will be averaged over all values of the vertical coordinate, $y_0$. The same procedure applied along
the $y$ direction gives $G_y(y)$. From the exact relation between $G_x$ and $S_x$,\cite{us_theor}
\begin{equation}
G_x(x) = \dfrac{2}{\pi}\int_0^\infty dk_x\, \left[1-\cos( k_x x)\right] S_x(k_x),
\end{equation}
we get the scaling law for $G_x$ as a function of the roughness exponent defined in Fourier space
\begin{equation}
G_x(x) \sim x^{2\tilde{\alpha}_x -\zeta -1}.
\end{equation}
Finally, by equating this last result with the scaling Ansatz for $G_x$ in real space, we are able to
relate the different roughness exponents
\begin{equation}
\label{eq:rough_x}
2\alpha_x = 2\tilde{\alpha}_x - \zeta -1,
\end{equation}
and, by applying the same reasoning to the $y$ direction, we get the relation
\begin{equation}
\label{eq:rough_y}
2\alpha_y = 2\tilde{\alpha}_y - 1/\zeta -1.
\end{equation}
Equations  \eqref{eq:rough_x} and \eqref{eq:rough_y} immediately convert Eqs.\ \eqref{eq:psd1d_scaling} and \eqref{eq:psd1d_scaling2} into \eqref{eq:psd1d_scaling_b} and \eqref{eq:psd1d_scaling_b2}. Moreover, it is simple to show that, in fact,
\begin{equation}
\zeta = \dfrac{\tilde{\alpha}_x}{\tilde{\alpha}_y} =  \dfrac{\alpha_x}{\alpha_y}.
\end{equation}

The anisotropic scaling Ansatz expressed by Eq.\ \eqref{psd2d} can be shown to be equivalent to others
that are known in the literature on critical systems, see, e.g.,  Refs.\ \onlinecite{pastor-satorras:1998} and \onlinecite{schmittmann:2006}, and references therein. Actually, it is an analog for nonconserved noise
fluctuations of the behavior found in so-called driven-diffusive systems.\cite{schmittmann:book} Nevertheless, an advantage of the present formulation is that one can relate the scaling behaviors of different 2D and 1D correlation functions with one another in a way that readily generalizes the isotropic behavior. In practice, this is a convenient consistency check that enables safe conclusions on the occurrence of the interesting property of SA in a given experiment, once compatibility ensues between the expected scaling behavior of the various observables as measured on the same set of data. We are not aware of such a type of analysis in the literature (see a related analysis of SA behavior in Ref.\ \onlinecite{keller:2009}), so that the next section is devoted to the discussion of a couple of experimental systems in which SA scaling, as described in the present section, is indeed shown to occur.

Note also that Eq.\ \eqref{psd2d} has been postulated once the PSD function $S(\mathbf{k})$ has reached a steady state
and is time independent.
The experimental data we report about in the next section satisfy this condition.
Nevertheless, in order for this type of behavior to properly become an anisotropic generalization of the so-called Family-Vicsek (FV) Ansatz that applies to isotropic kinetic roughening,\cite{barabasi:book} time dependence should be allowed for. The FV Ansatz is typically formulated in terms of the short and long time behavior for the surface roughness $W^2(t) = \langle (h-\bar{h})^2 \rangle = \int S(\mathbf{k}) \; {\rm d}\mathbf{k}$. Thus,\cite{barabasi:book} $W \sim t^{\beta}$ for $t \ll t^{1/z}$, while $W \sim L^{\alpha}$ for $t \gg t^{1/z}$, where $z$ is an independent exponent, $t^{1/z}$ is an estimate of a length scale below which nontrivial correlations have built up among height values at different substrate positions, and $\beta = \alpha/z$. In the SA case, it is natural to consider corresponding roughness functions $W_{x,y}$ for one-dimensional cuts of the surface, so that one would expect $W_{x,y} \sim t^{\beta_{x,y}}$ for $t \ll t^{1/z_{x,y}}$, while $W_{x,y} \sim L^{\alpha_{x,y}}$ for $t \gg t^{1/z_{x,y}}$. This is indeed the behavior compatible with the steady state Eq.\ \eqref{psd2d}, for which one can show \cite{us_theor} that moreover $z_y = z_x/\zeta$, and thus $\beta_x = \alpha_x/z_x = \beta_y = \alpha_y/z_y$. Hence, overall there are only three independent critical exponents characterizing a time-dependent SA surface, e.g., $\alpha_x$, $\zeta$, and $z_x$, from which all other exponents
described in this section can be obtained.

\section{Experiments}
\label{sec:exp}

In this section we show that the previous scaling analysis does provide a consistent description
of the anisotropic kinetic roughening properties of actual experimental surfaces. In particular, we choose
to produce the latter through erosion of silicon targets by IBS at low-intermediate ion energies.\cite{munoz-garcia:2009,chan:2007} By tuning experimental conditions such as the angle of incidence
of the ion beam onto the surface, average ion energy, etc., it is possible to obtain surfaces with varying
topographical properties, from disordered and rough to nanopatterned. In our case, these properties will
correlate with the simultaneous codeposition of metallic impurities, or the absence thereof, as recently
shown.\cite{redondo-cubero:2012,macko:2011} At any rate and as mentioned above, even in the morphologically unstable
case, it may be possible to find a range of scales within which scale invariance (i.e., kinetic roughening) occurs;
that is the one we will address. This is the case for large classes of pattern-forming systems.\cite{nicoli:2010}



In each section below, the analysis of the experimental data, and the verification of the scaling Ansatz presented in Sec.\ \ref{sec:asa}, has been carried out following the next several steps:
\begin{itemize}
\item For each experimental condition the irradiation dose guarantees that the surface has reached a steady state (as reflected in the time independence of the correlation functions measured).
\item A single sample is scanned (by STM or AFM) using different resolutions, and over regions of different sizes, which we will call windows. These scans provide us with different images, encoded in matrices, of the same sample.
\item For each matrix, we compute two quantities: the two-dimensional  PSD $S(k_x,k_y)$, and the PSDs of one-dimensional  cuts of the surface, $S_{x,y}(k_{x,y})$. The 2D PSD is then averaged over all the images we have. For the 1D PSDs, as explained in Sec.\ \ref{sec:asa}, we also perform averages over all the lines and columns of each matrix, respectively.
\item Next, we consider the projections $S(k_x,0)$ and $S(0,k_y)$ of the two-dimensional PSD onto the $k_x$ and
$k_y$ axes. According to our Ansatz, these profiles should scale as power laws of $k_{x,y}$ with exponents $2 \tilde \alpha_x$ and $2\tilde \alpha_y$, respectively, that are estimated by fits over appropriate scaling ranges.
\item Using formulas \eqref{eq:rough_x} and \eqref{eq:rough_y}, we compute real-space roughness exponents $\alpha_{x,y}$ and, through equations \eqref{eq:psd1d_scaling_b} and \eqref{eq:psd1d_scaling_b2}, we predict the scaling behavior for the one-dimensional PSDs.
\item Finally, we verify if the one-dimensional PSDs indeed scale in a form that is consistent with the two-dimensional PSD, as required by our scaling theory.
\end{itemize}

\subsection{Morphologically stable case}


For the morphologically stable condition that we will study, the experiments were performed in a scanning tunneling
microscopy (STM) apparatus with a base pressure below $1 \times 10^{-10}$ mbar.\cite{macko:2011}
The system is equipped with a differentially pumped fine focus ion source. The fine focus ion beam exposed only the sample and thus impurity effects were avoided. A Si(100) sample was irradiated at room temperature with 2 keV
Kr$^+$ ions at an ion incidence angle $\theta = 81^{\circ}$ with the surface normal. The reproducibility of the angle is better than $0.5^{\circ}$, the error of the angle's absolute value is $1^{\circ}$. The sample was exposed to an ion fluence of $2 \times 10^{22}$ ions m$^{-2}$, with an average ion flux of $3 \times 10^{17}$ ions m$^{-2}$ s$^{-1}$. The flux is specified here for the sample plane. It was controlled by a Faraday cup movable to the sample position. After ion exposure, the samples were imaged {\em in situ} by STM.

\begin{figure}[t!]
\epsfig{clip=,width=0.45\textwidth,file = 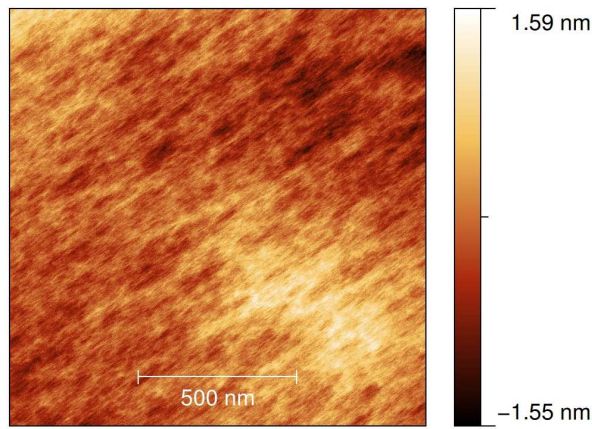}
\caption{STM top views of the surface morphology for the 
contamination-free sample. Image size is $1.3 \times 1.3$ $\mu$m$^2$. 
}
\label{fig:morph_81_83deg}
\end{figure}

A representative image is shown in Fig.\ \ref{fig:morph_81_83deg}.
Superficial naked-eye inspection already suggests the existence of a long-wave corrugation which does not seem
particularly ordered but does have a preferred orientation, that coincides with the projection of the ion-beam
onto the substrate plane. Actually, this is taken into account in the analysis of the topography.
Specifically, 
the ion beam comes from the lower left corner with an angle of 35$^\circ$ to the $x$ axis.
After rotating the images so that the ion-beam projection coincides with the $x$ (horizontal) axis, rectangular regions have been cut out from the original figures and provide the data for our analysis, see Ref.\ \onlinecite{suppl}.

\begin{figure}[t!]
\epsfig{clip=,width=0.5\textwidth,file = 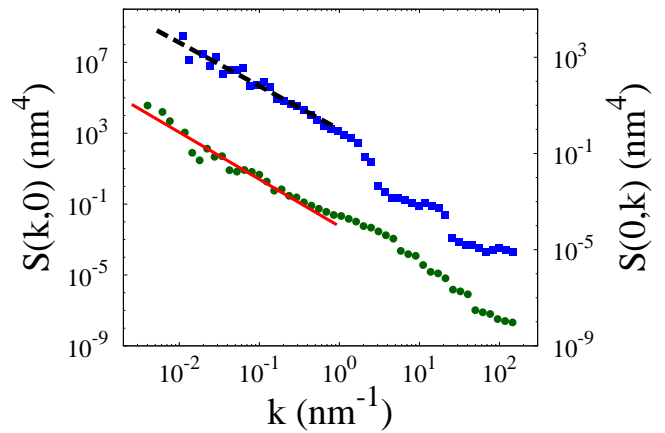}
\caption{One-dimensional projections $S(k,0)$ (green circles, left axis) and $S(0,k)$ (blue squares, right axis) of the two-dimensional PSD for the contamination-free sample, averaged over eight  windows. The solid red and dashed black lines are  fits obtained by least squares, the resulting exponents being $2\tilde{\alpha}_x \simeq 2.66$ and $2\tilde{\alpha}_y \simeq 1.80$. These lines correspond to the exact stationary behavior of Eq.\ \eqref{eq_n-m}
for $\nu_x = 178$, \mbox{$\nu_y = 0.99$}, and $D=1$ with nm and s for space and time units, respectively.}
\label{fig:michely_81deg_meanpsd2d}
\end{figure}

Projections of the two dimensional PSD along the $k_x$ and $k_y$ axes are presented in Fig.\ \ref{fig:michely_81deg_meanpsd2d}. 
These plots have been obtained after averaging the $S(\mathbf{k})$ functions obtained for different window sizes, as detailed in the Supplemental Material, to improve statistics.\cite{suppl}
Note that power-law behavior takes place for length scales that are larger than approximately 6 nm, while for shorter distances (larger corresponding values of $k$) different behavior is obtained. Thus, we will restrict our analysis in this case to the small $k$ range, thus indeed facing the properties of the long-wavelength corrugation just mentioned.

\begin{figure}[t!]
\epsfig{clip=,width=0.5\textwidth,file = 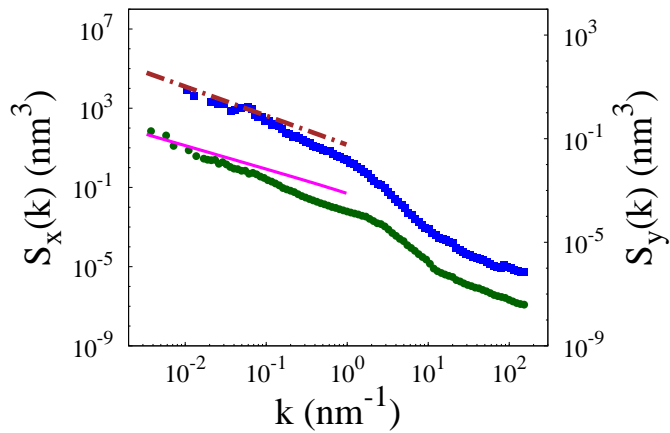}
\caption{PSD of one-dimensional cuts $S_x(k)$ (green circles, left axis) and $S_y(k)$
(blue squares, right axis) of the $\theta=81^{\circ}$ sample, averaged over eight windows. The solid magenta line and the dot-dashed brown line correspond to Eqs.\ \eqref{eq:psd1d_scaling_b} and \eqref{eq:psd1d_scaling_b2}, respectively, with values given by Eq.\ \eqref{eq:Michely_81deg_exponents_real}, and are {\em not} fits to the data. Rather, they have been computed using Eq.\ \eqref{psd1d_finite} for Eq.\ \eqref{eq_n-m} with $\nu_x = 178$, $\nu_y = 0.99$, and $D=1$, with nm and s for space and time units, respectively.}
\label{fig:michely_81deg_meanpsd1d}
\end{figure}

By performing a fit over the small wave-vector region in Fig.\ \ref{fig:michely_81deg_meanpsd2d}, we estimate the values of the momentum-space roughness exponents to be
\begin{equation}
\begin{split}
  2\tilde \alpha_x = 2.66 \pm 0.02, &\qquad \qquad 2\tilde \alpha_y = 1.80 \pm 0.02,   \\[5pt]
 & \hskip -10pt \zeta = 1.48 \pm 0.02, 
\end{split}
\label{eq:Michely_81deg_exponents_momentum}
\end{equation}
where error bars come from statistical uncertainty in the fits. Hence, strong anisotropy occurs, $\zeta\neq 1$. The corresponding exponents characterizing the power-law decays of the 1D PSD functions should be
\begin{equation}
\label{eq:Michely_81deg_exponents_real}
2\alpha_x + 1 = 1.18 \pm 0.02, \qquad 2\alpha_y + 1 = 1.12 \pm 0.02 .
\end{equation}
Using these values in Eqs.\ \eqref{eq:psd1d_scaling_b} and \eqref{eq:psd1d_scaling_b2}, the scaling behavior
of $S_{y}(k_{y})$ extracted from our Ansatz agrees well with experimental data as shown in Fig.\ \ref{fig:michely_81deg_meanpsd1d}, while such an agreement is not reached in the case of $S_x(k_x)$.
In order to understand this disagreement, note first that in the present system both real-space exponents are quite close
to being effectively zero, in view of the statistical fluctuations in the data. In the context of critical phenomena and power-law behavior this would be associated with logarithmic (rather than proper power-law) behavior for the real space correlation functions.\cite{barabasi:book,henkel:book_v1} Incidentally, when issues like this arise, the assessment of kinetic roughening behavior is usually more clear cut for Fourier-space observables than for real-space correlation functions, see Ref.\ \onlinecite{cuerno:2004} and references therein. On the other hand, as we can see in Fig.\ \ref{fig:michely_81deg_meanpsd1d}, the scaling behavior predicted by the simple power law $S_{x}(k_{x}) \sim k_{x}^{-(2\alpha_{x}+1)}$ has not been reached by the experimental data, especially for the smallest wave-vector values. An analogous inconsistency between the scaling behavior of the two-dimensional PSD and that of the 1D PSDs has also been observed in continuum models of SA surfaces.\cite{us_theor} The key to understanding the discrepancy is the fact that, for the small numerical values obtained for $\tilde{\alpha}_{x,y}$, integrals such as Eq.\ \eqref{psd1d} converge slowly to their asymptotic properties, in the sense that non-negligible corrections occur to the leading terms $k^{-(2\alpha_{x,y}+1)}$ in a small $k_{x,y}$ expansion. In other words, one needs to rather compute the full integrals \eqref{psd1d} that, for a finite space resolution, read
\begin{equation}
\label{psd1d_finite}
S_{x,y}(k_{x,y}) \propto \int_0^{\pi/a} dk_{y,x} \, S(k_x,k_y) ,
\end{equation}
where $a$ is the smallest length scale in the region explored. Even in such a case, convergence to the asymptotic behavior may occur only at length scales that may not be easily accessible in an experiment.

In the present case, based on results from Ref.\ \onlinecite{us_theor}, we can write down a continuum model with the same scaling exponents as measured in \eqref{eq:Michely_81deg_exponents_momentum}, for which SA behavior holds rigorously, and analyze for it the behavior of the 1D PSD functions. Thus, as a proxy for the scaling behavior of the present experimental data, we can write down the following {\em linear} evolution equation for the Fourier components of the height field, $\tilde{h}(\mathbf{k},t)$,
\begin{equation}
\label{eq_n-m}
\partial_t \tilde{h} = -(\nu_x |k_x|^{2.66} + \nu_y |k_y|^{1.80}) \tilde{h} + \tilde{\eta} .
\end{equation}
Note, this equation has a similar structure to the one Eq.\ \eqref{eq_2-4} would have in Fourier space, only with
values of the exponents that are those from Eq.\ \eqref{eq:Michely_81deg_exponents_momentum}, instead of $2\tilde\alpha_x = 2$ and $2\tilde\alpha_y = 4$, as would be the case for the latter. Equation \eqref{eq_n-m} has a complicated expression in real space, and can be interpreted as a model of anisotropic {\em nonlocal} surface relaxation.\cite{nicoli:2009} It can be readily shown\cite{us_theor} that, in the long time limit, Eq.\ \eqref{eq_n-m} follows the scaling Ansatz \eqref{psd2d} {\em exactly} with exponents given by \eqref{eq:Michely_81deg_exponents_momentum}. Actually, the power-law fits in Fig.\ \ref{fig:michely_81deg_meanpsd2d} are the exact stationary behavior of Eq.\ \eqref{eq_n-m}, provided we use $\nu_x = 178$, $\nu_y = 0.99$, and $D=1$, with nm and s for space and time units, respectively. Once this is the case, the behavior of the 1D PSD functions for Eq.\ \eqref{eq_n-m} follows without the possibility to further fits, and is shown in Fig.\ \ref{fig:michely_81deg_meanpsd1d}. As we see, there is good quantitative agreement with experimental data for $S_y(k_y)$, while in the case of $S_x(k_x)$ the experimental data overshoot the theoretical curve at small $k_x$ values. For a full class of models related with Eq.\ \eqref{eq_n-m} this is an indication that the true asymptotic behavior is reached at still larger distances,\cite{us_theor} and we believe this is the case for the present set of data. Note, we are not claiming that Eq.\ \eqref{eq_n-m} provides the description of the full dynamics of the latter. We do believe it shares with it the same asymptotic behavior, and this is possible even if the ``true'' dynamical equation for this system is a different, even nonlinear, one.\cite{us_theor} An interesting physical conclusion for the modeling of IBS problems stems from the fact that Eq.\ \eqref{eq_n-m} preserves the average value of the height field along the system evolution, since $\langle \tilde{h}(\mathbf{0},t) \rangle = \langle \bar{h}(t) \rangle = \mbox{const}$. This suggests in particular that erosive mechanisms, \cite{chan:2007,munoz-garcia:2009} that do not preserve this quantity, are (numerically) negligible under the present experimental conditions, which agrees with recent experimental results for similar systems, see, e.g., Ref.\ \onlinecite{madi:2011} and references therein.

\subsection{Morphologically unstable case}

\begin{figure}[t!]
\epsfig{clip=,scale = 0.35, file = 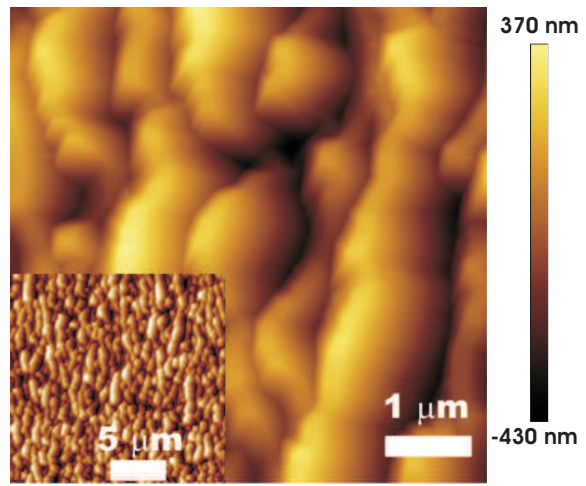}
\caption{AFM top view of the morphology of Si(100) irradiated for 8 h with $40$ keV Ar$^+$ ions and concurrent Fe codeposition. Image size is $6 \times 6 \; \mu$m$^2$.}
\label{fig:morph_luis}
\end{figure}


The irradiation experiments in the morphologically unstable case were performed with a $40$ keV Ar$^+$ beam extracted
from a Danfysik $1090$ ion implanter with a base pressure of $5 \times 10^6$ mbar, similar to Ref.\ \onlinecite{redondo-cubero:2012}. The ions impinged on the single-crystal Si($100$) targets ($1 \times 1 \text{ cm}^2$) at $60^{\circ} \pm 5^{\circ}$ with respect to the surface normal with a current density of $18 \, \mu \text{A} \text{ cm}^{-2} $ in the sample plane. A steel plate ($1.5 \; \text{mm}$ high) placed adjacent to the Si target acted simultaneously as Fe source and sample holder. To obtain homogeneous irradiation, the focused beam was scanned with a magnetic $x-y$ sweeping system in such way that both the Si surface and the steel target were bombarded. The irradiation time was eight hours. The resulting surface morphology was imaged {\em ex-situ} by AFM operating in the dynamic mode with  Nanoscope IIIa equipment (Veeco$^\copyright$). Silicon cantilevers, with a nominal radius $r$ of $8 \text{nm}$ and opening angle smaller than $52^{\circ}$, were employed. A representative topography is shown in Fig.\ \ref{fig:morph_luis}. The projection of the ion
beam coincides with the $x$ (horizontal) axis in the figure.

\begin{figure}[t!]
\epsfig{clip=,width=0.5\textwidth,file = 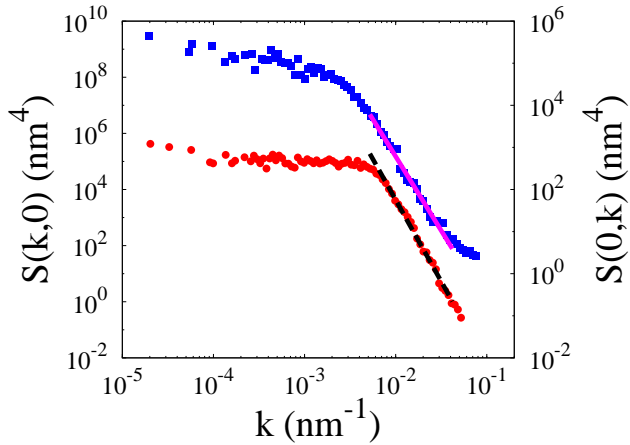}
\caption{One-dimensional projections $S(k,0)$ (red circles, left axis) and $S(0,k)$ (blue squares, right axis) of the two-dimensional PSD of the samples illustrated in Fig.\ \ref{fig:morph_luis}, after an average over seven  windows. The black dashed line and the solid magenta line are fits obtained by least squares, the resulting exponents being $2\tilde{\alpha}_x \simeq 5.81$ and $2\tilde{\alpha}_y \simeq 3.74$.}
\label{fig:luis_meanpsd2d}
\end{figure}
\begin{figure}[t!]
\epsfig{clip=,width=0.5\textwidth,file = 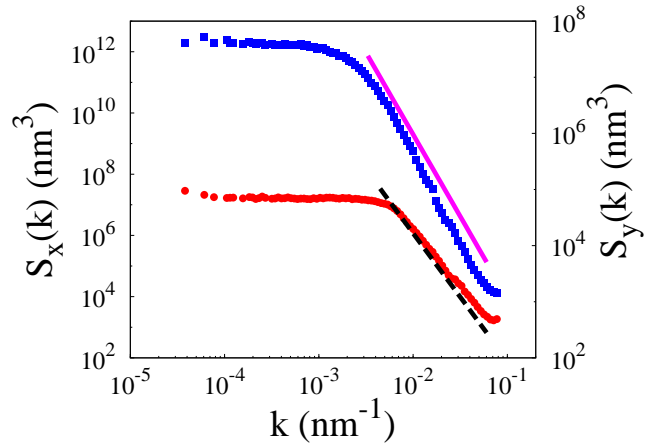}
\caption{PSD of one-dimensional cuts $S_x(k)$ (red circles, left axis) and $S_y(k)$
(blue squares, right axis) of the samples illustrated in Fig.\ \ref{fig:morph_luis}, after an average over seven windows. The black dashed line and the solid magenta line correspond to Eqs.\ \eqref{eq:psd1d_scaling_b} and \eqref{eq:psd1d_scaling_b2}, respectively, with exponents values given by Eq.\ \eqref{eq:Luis_exponents_real}.}
\label{fig:luis_meanpsd1d}
\end{figure}

Although from the physical point of view the present system is perhaps more complex (due to the not well understood role of impurities in the nanopatterning process), the occurrence of SA seems in principle clear cut. We employ the same procedure as above. After computing the two-dimensional PSD, we show its two projections $S(k_x,0)$ and $S(0,k_y)$ in Fig.\ \ref{fig:luis_meanpsd2d}. Both plots appearing in the latter figure immediately suggest the existence of a characteristic length scale around 800-900 nm, above which no proper scaling takes place. We associate this scale with the wavelength of the pattern that develops under these experimental conditions.
Nevertheless, at smaller scales (larger $k$ values) power-law behavior appears that is different in the
$k_x$ and $k_y$ directions, as described by exponent values
\begin{equation}
\begin{split}
  2\tilde \alpha_x = 5.81 \pm 0.06,& \qquad\qquad 2\tilde \alpha_y = 3.74 \pm 0.2,  \\[5pt]
 &\hskip -10pt \zeta = 1.55 \pm 0.01, 
\end{split}
\label{eq:Luis_exponents_momentum}
\end{equation}
that are obtained by least squares. Note, $\zeta\neq 1$.
Thus, according to Eqs.\ \eqref{eq:rough_x} and \eqref{eq:rough_y}, exponents characterizing the power-law decay of the 1D PSD functions should be
\begin{equation}
2\alpha_x +1 = 4.26 \pm 0.14, \qquad 2\alpha_y + 1 = 3.10 \pm 0.2 .
\label{eq:Luis_exponents_real}
\end{equation}
Using these values in Eqs.\ \eqref{eq:psd1d_scaling_b} and \eqref{eq:psd1d_scaling_b2}, the agreement with the experimental data obtained for the one-dimensional PSDs for the corresponding range in $k$ is very good, as shown in Fig.\ \ref{fig:luis_meanpsd1d}.

The appearance of scaling behavior at length scales below the one associated with the morphological instability
suggests as a cause the influence of the average shape associated with it, rather than kinetic roughening, similar to the case of mound formation in epitaxial growth on high symmetry surfaces.\cite{michely:book} In order to explore this
possibility, we have generated artificial topographies made up by an array of semi-ellipsoids that are placed off their
ordered positions by a random error, or a disordered array of faceted features.\cite{suppl} In both cases, while the interplay between disorder and the smooth shape of the basic motif (semi-ellipsoid or sawtooth) does give rise to power law behavior of 2D and 1D PSD functions below the corresponding characteristic sizes, the scaling exponents do {\em not} fulfill the relations expected within our scaling Ansatz. Rather, scaling in these cases seems related with the smooth geometry of the basic pattern,\cite{keblinski:1993} combined with fluctuations in its space arrangement.

We thus suggest strongly anisotropic kinetic roughening at submicron scales (above which scale invariance is lost) to account for the present experimental system features. To some extent, such a behavior is complementary to the one that is typical of e.g.\ the KS system,\cite{munoz-garcia:2009} in which a small-scale pattern becomes disordered at sufficiently large scales at which kinetic roughening occurs.\cite{nicoli:2010} Although in terms of the latter class of systems it may look somewhat peculiar, the behavior found for the experiments just analyzed is again readily obtained in appropriate simple model systems. Thus, one can generalize, e.g., Eq.\ \eqref{eq_2-4} to
\begin{equation}
\label{eq_2-4_damped}
\partial_t h = -\mu h + \nu_x \partial_x^2 h - \nu_y \partial_y^4 h + \eta(x,y,t),
\end{equation}
where the linear term with coefficient $\mu >0$ accounts for a physical mechanism (e.g., a wetting layer) that favors
a specific value of the height (say, $h=0$). For Eq.\ \eqref{eq_2-4_damped}, at $t\to\infty$ one can exactly obtain $S(k_x, k_y) \sim (\mu + \nu_x k_x^2 + \nu_y k_y^4)^{-1}$. This implies that $S(k_x, k_y)$ does {\em not} scale with $k_{x,y}$
(but, rather, becomes a constant) for the largest length scales (smallest $k_{x,y}$ values), thus breaking
scale invariance at such large scales. However, kinetic roughening precisely as in Eq.\ \eqref{psd2d} does still occur for (large) $k_{x,y}$ values whose contribution to $S(k_x, k_y)$ dominates over $\mu$. While we are not saying that Eq.\ \eqref{eq_2-4_damped} describes the experiments just analyzed, we believe the latter may correspond to a situation that is qualitatively captured by this example. Currently there is a large effort to pursue  a more quantitative description of IBS experiments in the presence of metallic contamination that takes into account these type of effects, see Ref.\ \onlinecite{bradley:2011} and references therein.

\section{Discussion and conclusions}
\label{sec:disc}

We have verified a scaling Ansatz for strongly anisotropic kinetic roughening properties for two
experimental systems. First of all, our results provide a consistent experimental assessment of such type of
behavior, which calls for the development of theoretical models that can describe in detail the experimental properties
observed. This would add to an improved knowledge on the phenomena driving the specific systems considered,
as well as to our understanding of driven systems in general. From a more practical point of view, our work illustrates several forms of occurrence of SA, both in the absence and presence of morphological instabilities, and at (sufficiently large) scales that remain either above or below a privileged (pattern) size.

Generally speaking, even in the  absence of typical length scales, it is important to keep in mind that our scaling Ansatz applies only in a hydrodynamic limit. Therefore, in actual physical systems, effects due to finite-size and/or finite-space resolution of measurement techniques could hinder a clear-cut scaling of the correlation functions, thus preventing the observation of a consistent behavior between the one and two-dimensional PSDs. As we have seen, this ``slow convergence'' issue appears to be more pronounced for integrated quantities, such as the one-dimensional PSDs
[see formula \eqref{psd1d}], and has been studied in more detail in Ref.\ \onlinecite{us_theor} by means of analytical arguments as well as numerical simulations. Thus, the two-dimensional PSD, conveniently averaged over several samples (or windows of the same sample), appears to be a more reliable observable in order to assess anisotropic scaling. However, in experiments it is seldom possible to have a sufficient number of samples over which to perform such an average, and the two-dimensional PSD frequently appears too noisy to extract reliable values of the exponents. The one-dimensional PSDs have a higher signal-to-noise ratio due to the fact that, for each window, they are averaged over all the rows or columns of the
matrix. For this reason, despite the fact that they are more sensitive to finite size effects, such quantities
do indeed provide relevant information about surface features.

We can hypothesize that this type of complication may have hindered a more frequent observation of anisotropic scaling in the literature. We hope that, once the analysis has been clarified, the identification of this challenging type of behavior becomes simplified and we can thus understand it better. Perhaps an analogy can be drawn with the case of anomalous scaling. This is a type of kinetic roughening behavior that was usually associated with large values of the roughness exponent, that induced many wrong assessments of scaling behavior both in experiments and models.\cite{cuerno:2004} Incidentally, such ``anomalous'' values usually introduced slow convergence properties in correlation functions. Once anomalous scaling was identified and systematized,\cite{lopez:97,ramasco:00,lopez:2005} it has been indeed assessed in different thin-film experimental systems, from, e.g., electrochemical deposition\cite{schwarzacher:2004} to chemical vapor deposition,\cite{yanguas-gil:2006} metal dissolution,\cite{cordoba-torres:2009} etc. We can only hope for a parallel development regarding strongly anisotropic surface kinetic roughening in the near future.

\vspace*{0.2cm}

\begin{acknowledgments}
We are very much indebted to A.\ Keller for discussions and exchange. 
We gratefully acknowledge the support by Deutsche Forschungsgemeinschaft through Forschergruppe 845.
Partial support for this work has been provided by MICINN (Spain) Grants No.\ FIS2009-12964-C05-01 and FIS2009-12964-C05-04. 
E.\ V.\ acknowledges support by Universidad Carlos III de Madrid.
\end{acknowledgments}

\end{document}